\documentclass[aps,twocolumn,superscriptaddress,showpacs,floatfix]{revtex4}
\usepackage{epsfig}
\input epsf

\begin{document}

\title{Free energy and configurational entropy of liquid silica:
fragile-to-strong crossover and polyamorphism}

\author{Ivan Saika-Voivod}
\affiliation{Department of Applied Mathematics,
University of Western Ontario, London, Ontario N6A~5B7, Canada}
\affiliation{Dipartimento di Fisica and Istituto Nazionale per la
Fisica della Materia, Universita' di Roma La Sapienza, Piazzale Aldo
Moro~2, I-00185, Roma, Italy}

\author{Francesco Sciortino}
\affiliation{Dipartimento di Fisica and Istituto Nazionale per la
Fisica della Materia, Universita' di Roma La Sapienza, Piazzale Aldo
Moro~2, I-00185, Roma, Italy}
\author{Peter H. Poole} 
\affiliation{Department of Applied
Mathematics, University of Western Ontario, London, Ontario N6A~5B7,
Canada} 
\affiliation{Department of Physics, St. Francis Xavier
University, Antigonish, Nova Scotia B2G 2W5, Canada}

\date{\today}

\begin{abstract}
Recent molecular dynamics (MD) simulations of liquid silica, using the
``BKS'' model [Van Beest, Kramer and van Santen, Phys. Rev. Lett. {\bf
64}, 1955 (1990)], have demonstrated that the liquid undergoes a
dynamical crossover from super-Arrhenius, or ``fragile'' behavior, to
Arrhenius, or ``strong'' behavior, as temperature $T$ is
decreased. From extensive MD simulations, we show that this
fragile-to-strong crossover (FSC) can be connected to changes in the
properties of the potential energy landscape, or surface (PES), of the
liquid.  To achieve this, we use thermodynamic integration to evaluate
the absolute free energy of the liquid over a wide range of density
and $T$.  We use this free energy data, along with the concept of
``inherent structures'' of the PES, to evaluate the absolute
configurational entropy $S_c$ of the liquid.  We find that the
temperature dependence of the diffusion coefficient and of $S_c$ are
consistent with the prediction of Adam and Gibbs, including in the
region where we observe the FSC to occur.  We find that the FSC is
related to a change in the properties of the PES explored by the
liquid, specifically an inflection in the $T$ dependence of the
average inherent structure energy.  In addition, we find that the high
$T$ behavior of $S_c$ suggests that the liquid entropy might approach
zero at finite $T$, behavior associated with the so-called Kauzmann
paradox.  However, we find that the change in the PES that underlies
the FSC is associated with a change in the $T$ dependence of $S_c$
that elucidates how the Kauzmann paradox is avoided in this system.
Finally, we also explore the relation of the observed PES changes to
the recently discussed possibility that BKS silica exhibits a
liquid-liquid phase transition, a behavior that has been proposed to
underlie the observed polyamorphism of amorphous solid silica.
\end{abstract}


\maketitle
\section{Introduction}

Liquid silica is the archetypal ``strong liquid,'' that is, a liquid
whose viscosity $\eta$ and other measures of relaxation follow closely
an Arrhenius behavior, $\log \eta \sim 1/T$~\cite{A91,richet}, where
$T$ is the temperature.  For most liquids, $\eta$ increases
significantly faster than an Arrhenius law as $T$ approaches the glass
transition temperature $T_g$; these liquids are referred to as
``fragile.''

Strong liquids such as silica are important as glass-forming systems.
In a strong liquid, $\eta$ varies less rapidly with $T$ near $T_g$,
compared to a fragile liquid.  As a consequence, a strong liquid can
be held in a desired range of $\eta$ over a wider range of $T$ than a
fragile liquid.  As every glassblower knows, this makes silica-based
systems easier to manipulate just above $T_g$ than any other commonly
available liquid.

The fundamental origins of strong behavior in glass-forming liquids is
also a subject of continuing interest.  We note in particular two
recent developments.  First, computer simulation work of Horbach and
Kob~\cite{HK99} using the ``BKS'' model of silica~\cite{BKS903} has
demonstrated that at high $T$, the model liquid exhibits fragile
behavior, and then crosses over to a regime of strong behavior upon
cooling.  The work of La Nave and coworkers, based on instantaneous
normal mode analysis, has shown that such a crossover is connected to
a progressive reduction in the number of diffusive directions in phase
space accessed by the system~\cite{selle}.  Such a ``fragile-to-strong
crossover'' (FSC) may be a general mechanism underlying the emergence
of strong behavior, and has since been studied for a number of
systems~\cite{jagla}.  We note that a crossover from a super-Arrhenius
to Arrhenius dynamics may be a general feature of liquids around the
so-called mode-coupling temperature~\cite{goetzepisa}, as is appearing
to emerge in recent numerical studies, thanks to the the larger
dynamical window made available by current computational
power~\cite{rinaldipre,MOSSA02,sastryjpcm}. However, the $T$ region
where equilibrium simulations can be performed is still limited, and
does not allow for a precise statement of the $T$-dependence below the
crossover temperature, as required to make final contact with models
for the glass transition~\cite{parisimezar,chandler, wolynes}.

Second, a growing body of computer simulation research has established
the importance of the potential energy landscape or surface (PES) for
understanding the dynamics of liquids near
$T_g$~\cite{goldstein,sw,S95,SDS98,SKT99,BH99,SCALA00,S01,caax,ds01,starr,MOSSA02,lanavejpcm}. The
PES refers specifically to $\cal U$, the instantaneous potential
energy hypersurface of the system, expressed as a function of the $3N$
coordinates ${q_i}$ that specify the positions of the $N$ atoms of the
system; i.e. ${\cal U}={\cal U}(q_1,q_2,...,q_{3N})$.  The properties
and topology of the PES have been carefully studied in the above cited
works, predominantly in the case of fragile liquids, resulting in
important insights into the equilibrium~\cite{eosprl} and
out-of-equilibrium~\cite{stprl,mossajpcm} thermodynamics of
supercooled states, and the connection between thermodynamics and
transport propertie~\cite{heuerjpcm,S01}.  However, the
relationship of the PES to the dynamic properties of strong liquids is
less well understood.  In this paper, our focus is to clarify this
relationship, and in particular, to determine if the FSC proposed for
liquid silica can be connected to properties of the PES.

Following previous studies of fragile liquids, our approach is to
apply the ``inherent structure'' formalism of Stillinger and
Weber~\cite{sw} to molecular dynamics (MD) computer simulation data
obtained for the BKS model of liquid silica.  In this approach, the
PES is partitioned into basins associated with the local minima of
$\cal~U$~\cite{goldstein,sw,S95}. Each minimum corresponds to a
particular configuration of atoms and is called an inherent structure
(IS). We denote by $e_{IS}$ the average potential energy of the IS's
associated with the basins sampled by the equilibrium liquid at a
given $T$ and volume $V$. An IS and its energy can be obtained in
computer simulation by carrying out a local minimization of $\cal U$
starting from an equilibrium liquid configuration.

As we will describe in detail below, the evaluation of $e_{IS}$,
combined with free energy calculations, allows us to calculate the
configurational entropy, $S_c$, of the
system~\cite{SKT99,S01,SCALA00,starr,MOSSA02}.  $S_c$ determines the
number of distinct configurations explored by the system, in this case
the basins of the PES.  In a liquid, diffusion is associated with the
exploration by the system of different basins of the PES.  The work of
Adam and Gibbs (AG) predicts a relationship (in the low $T$ limit)
between the characteristic relaxation time of the system and
$S_c$~\cite{AG65}.  The AG relation has been recently derived in a
novel way~\cite{wolynes}.  Generalizing the AG relation to the
diffusion coefficient $D$, the AG relation can be written as,
\begin{equation}
\label{eAG}
\frac{D}{T}= \mu_0 \exp \Biggl(-\frac{A}{TS_c}\Biggr),
\end{equation}
where $\mu_0$ and $A$ are presumed to be constant with respect to $T$.
In the context of liquid silica, an interesting test of the robustness
of the AG relation is possible by checking if Eq.~\ref{eAG} is obeyed
throughout the region in which the FSC occurs.  If so, the AG relation
then provides a basis for connecting transport behavior (quantified
by $D$) to the properties of the PES (quantified by $S_c$ and
$e_{IS}$).  

In a recent letter~\cite{NAT01}, we showed that liquid BKS silica
behaves in a manner that allows $S_c$ to be calculated from $e_{IS}$,
and that $D$ and $S_c$ are related as predicted by the AG relation.
We were thereby able to show that the FSC in liquid silica is
associated with a change in the $T$ dependence of $e_{IS}$, i.e. a
change in the nature of the PES explored by the system as $T$
decreases.  We also found that this observation in turn has
implications for other behavior observed in BKS silica, in particular,
the possible occurrence of a liquid-liquid transition, and the
behavior of the liquid as related to the so-called Kauzmann paradox.

To reach such conclusions, extensive MD simulations are required over
a wide range of $V$ and $T$, to calculate thermodynamic and transport
properties, as well as careful examination of the IS properties.  In
addition, the absolute free energy of the liquid must be evaluated.
In the present work, we provide a detailed description of the methods
used to obtain the results summarized in Ref.~\cite{NAT01}, and also
provide an expanded analysis and discussion of the results.  This work
is organized as follows.  In Section~II we describe our MD
simulations, including the interaction potential used.  Section~III
provides a detailed description of the techniques we use to evaluate
$e_{IS}$, $S_c$, and the total free energy of the liquid.  Section~IV
presents the results of these calculations and provides a discussion
of their implications.

\section{Molecular Dynamics Simulations}

We carry out MD simulations at constant $V$.  Most of our results are
for a system of 444 Si atoms and 888 O atoms.  A few simulations are
carried out with a reduced number of particles (333 Si and 666 O
atoms) in order to access longer physical times scales.  Our MD
simulation program is based on the ``MDCSPC2'' source
code~\cite{MDCSPC}.  We also reproduce a subset of our results using a
code we have written independently of MDCSPC2.  Note that all molar
quantities are reported here in moles of atoms.

Our model of atomic interactions in silica, denoted here as
$\Phi_{BKS}$, is based on the BKS potential, modified in two ways.
First, the BKS potential energy for both the Si-O and O-O interactions
diverges unphysically to negative infinity at sufficiently small
distances, allowing ``fusion'' events to occur during simulation of
high $T$ systems.  To prevent this, $\Phi_{BKS}$ consists of the
standard BKS potential plus a short range term given by
\begin{equation}
4\epsilon_{\mu\nu}\Biggl[ \biggl( \frac{\sigma_{\mu\nu}}{r_{ij}} \biggr) ^{30}
                     -\biggl( \frac{\sigma_{\mu\nu}}{r_{ij}} \biggr) ^{6}\Biggr]
\end{equation}
where $r_{ij}$ is the interatomic separation between an atom $i$ of
species $\mu$, and an atom $j$ of species $\nu$.  To choose the
parameters $\epsilon_{\mu\nu}$ and $\sigma_{\mu\nu}$ (see
Table~\ref{tPAR}) we first identify the value $r_{ij}=r^\ast_{ij}$ at
which the inflection of the standard BKS potential occurs, below which
the divergence to negative infinity develops.  The parameters are
chosen so that the new potential increases monotonically, and without
inflections, as $r_{ij}$ decreases for $r_{ij}<r^\ast_{ij}$; and so
that the difference between the new and the old potentials is small
for $r_{ij}>r^\ast_{ij}$.  Similar approaches have been used in other
works~\cite{GG96,VK}.

The second modification to the standard BKS potential included in
$\Phi_{BKS}$ relates to the treatment of longer range interactions.
As is common in implementations of the BKS potential, we calculate the
long range contributions to the Coulombic potential energy using the
Ewald summation technique, with the dipole surface term set to
zero~\cite{AT893}.  The reciprocal space summation is carried out to a
radius of six reciprocal lattice cell widths.  In this approach, the
real space contributions to the BKS potential are usually cut off
discontinuously at a specified distance, often chosen as $L/2$, where
$L$ is the length of an edge of the simulation cell.  However, we
study systems over a wide range of density $\rho$, and we desire a
potential for which the cut-off is independent of $L$.  Also, for
accurate determination of inherent structures, we wish to remove
discontinuities in the potential energy arising from cut-offs, and to
remove any $L$ dependence from long range corrections associated with
the cut-off.

To achieve these goals, instead of discontinuously cutting off the
real space potential contributions, we introduce a switching function.
At a fixed distance $R_s=0.77476$~nm the real space terms of the
standard BKS potential are replaced by a 5th degree polynomial that
tapers smoothly to zero over the range $R_s<r_{ij}<R_c$, where
$R_c=1$~nm.  The polynomial coefficients and the value of $R_s$ are
chosen so that the potential is continuous up to and including second
derivatives at both $r_{ij}=R_s$ and $r_{ij}=R_c$; and so that the
potential and its first two derivatives are monotonic for
$R_s<r_{ij}<R_c$, and go to zero as $r_{ij}\to R_c$.  These choices
depend on the Ewald parameter $\alpha$ that occurs in both the real
and reciprocal space contributions to the potential energy.  For all
$L$, we choose $\alpha = 2.5 \,{\rm nm}^{-1}$ to ensure sufficient
convergence of the potential energy in the reciprocal space summation
for the densities studied.  The value $R_c=1$~nm where the switching
function reaches zero is chosen to include third Si-Si neighbor
interactions at most densities studied.

The real space contribution to $\Phi_{BKS}$, denoted here as $\phi$,
is therefore a piece-wise defined function of the form,
\begin{eqnarray}\label{eqREAL}
\phi (r_{ij}\le R_s) &=& \frac{q_\mu q_\nu}{4 \pi \varepsilon}
                 \frac{{\rm{erfc}}(\alpha r_{ij})}{r_{ij}} \nonumber
                 \\ &+& A_{\mu\nu}e^{-B_{\mu\nu} r_{ij}} + \frac
                 {C_{\mu\nu}}{r_{ij}^6} \nonumber \\ &+&
                 4\epsilon_{\mu\nu}\Biggl[ \biggl(
                 \frac{\sigma_{\mu\nu}}{r_{ij}} \biggr) ^{30} -\biggl(
                 \frac{\sigma_{\mu\nu}}{r_{ij}} \biggr) ^{6}\Biggr]\\
                 \phi (R_s<r_{ij}<R_c) &=& D_{\mu\nu}(r_{ij} - R_s)^5
                 \nonumber \\ &+& E_{\mu\nu}(r_{ij} -R_s)^4 \nonumber \\
                 &+& F_{\mu\nu}(r_{ij} -R_s)^3 \\ \phi (r_{ij}\ge R_c)
                 &=& 0
\label{switch}
\end{eqnarray}
where $\rm{erfc}(x)$ is the complementary error function and
$\varepsilon$ is the permittivity constant.  The parameters are given
in Table~\ref{tPAR}.  

Note that the above modifications have the consequence that the
average potential energy $U$ and pressure $P$ obtained using
$\Phi_{BKS}$ differ from those obtained using the standard BKS
potential.  We find that the differences are approximately independent
of $T$ along isochores.  At $T=4000$~K and $\rho=2.3072$~g/cm$^3$, we
find that $\Phi_{BKS}$ gives a $P$ value $0.25$~GPa greater than the
standard BKS potential, and $U$ is $2.5$~kJ/mol higher.  At the same
$T$ and $\rho=3.8995$~g/cm$^3$, the respective differences are
$0.9$~GPa and $4.4$~kJ/mol higher.  These are not large differences on
the scale of our measurements, and the qualitative behavior of the
system is, as shown below, consistent with that found in other studies
based on the BKS model.

For the free energy calculation to be described below, we also perform
MD simulations using a binary Lennard-Jones (LJ) potential, in which
two atomic species (also labeled ``Si'' and ``O'') occur in the same
$1:2$ proportion as in SiO$_2$.  The LJ pair potential is of the form
\begin{equation}
\Phi_{LJ} = 4g_{\mu\nu}\Biggl[
           \biggl(\frac{s_{\mu\nu}}{r_{ij}}\biggr)^{12} -
           \biggl(\frac{s_{\mu\nu}}{r_{ij}}\biggr)^6 \Biggr] -
           \Phi^{\rm shift}_{\mu\nu}
\label{lj}
\end{equation}
The pair potential is cut off at $r_{ij}=2.5 s_{\mu\nu}$ and
$\Phi^{\rm shift}_{\mu\nu}$ is determined so that
$\Phi_{LJ}(r_{ij}=2.5s_{\mu\nu})=0$.  These potential parameters are
given in Table~\ref{tPAR}.

In order to obtain equilibrium properties we use the following
procedure.  We equilibrate the liquid using velocity rescaling for a
time $\tau$ long enough to allow Si atoms to diffuse an average of
$0.2$~nm, after significant relaxation of $P$ and $U$ have
disappeared from the system history.  The interval of velocity
rescaling varies from $10$ to $1000$ time steps depending on $T$.  The
time step for all runs is $1$~fs, except for $T=7000$~K, where the
time step is $0.5$~fs.  Velocity scaling is then turned off and the
system is evolved in a constant $(N,V,E)$ ensemble for at least $10
\tau$.  ($E$ is the total energy.)  Using this approach, there is no
appreciable drift in $E$ during the constant $(N,V,E)$ phase, over
which we calculate equilibrium quantities.

For the lowest $T$ where relaxation is slowest we modify this
procedure to improve our sampling of phase space: we conduct up to
five independent runs, with the constant $(N,V,E)$ phase of each run
lasting at least $2 \tau$.  The reported properties (including $T$)
are averages over both time and over the independent runs.  Thus
averages for low $T$ state points are also calculated over a total of
$10 \tau$, while at the same time the danger of an undetected trapping
in an out-of-equilibrium state is reduced through comparison of the
independent runs.

\begin{figure}
\hbox to\hsize{\epsfxsize=1.0\hsize\hfil\epsfbox{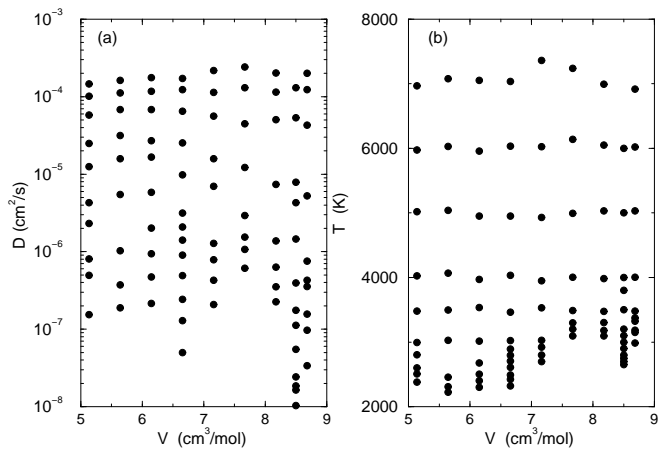}\hfil}
\caption{State points simulated in the (a) $V$-$D$ and (b) $V$-$T$
planes.  All points give results for equilibrated liquids.}
\label{points}
\end{figure}

The densities of the isochores simulated are given in
Table~\ref{tISO}, while the state points studied are shown in
Fig.~\ref{points}.  Note that we have studied the isochore at density
$2.3566 \,{\rm g/cm}^3$ in order to compare with previously published
work~\cite{HK99}.  The simulations along this isochore are those that
involve only $999$ atoms; all others model $1332$ atoms.

\begin{table}
\begin{tabular}{|l|c|c|c|}
\hline
$\mu$-$\nu$ & Si-Si  & Si-O & O-O \\
\hline
\multicolumn{4}{|c|}{standard BKS parameters} \\
\hline
$A_{\mu\nu}$ (J $\times 10^{-16}$)    & 0  & 28.845422 & 2.2250768  \\
$B_{\mu\nu}$  (nm$^{-1}$)      & 0  & 48.7318                & 27.6                   \\
$C_{\mu\nu}$  (J nm$^{6}$ $\times 10^{-23}$ )  & 0  &-2.1395327 &-2.8038308 \\
\hline
\multicolumn{4}{|c|}{$\Phi_{BKS}$: short range parameters} \\
\hline
$\epsilon_{\mu\nu}$ (J $\times 10^{-22}$)   & 0  & 4.963460     & 1.6839685    \\
$\sigma_{\mu\nu}$ (nm)   & 0  & 0.1313635              & 0.1779239              \\
\hline
\multicolumn{4}{|c|}{$\Phi_{BKS}$: switching function parameters} \\
\hline
$D_{\mu\nu}$  (J/nm$^5$ $\times 10^{-19}$) & -235.3529  & 122.0161  & -53.16278  \\
$E_{\mu\nu}$  (J/nm$^4$ $\times 10^{-19}$) & -117.7993  & 61.33742  & -26.25876  \\
$F_{\mu\nu}$  (J/nm$^3$ $\times 10^{-19}$) & -23.83785  & 12.33446  & -5.415203  \\
\hline
\multicolumn{4}{|c|}{$\Phi_{LJ}$ parameters} \\
\hline
$g_{\mu\nu}$  (kJ/mol)      & 23.0  & 32.0  & 23.0  \\
$s_{\mu\nu}$  (nm)          & 0.33  & 0.16  & 0.28  \\
\hline
\end{tabular}
\caption{Potential parameters used in this work for both $\Phi_{BKS}$
and $\Phi_{LJ}$.  Also required to specify $\Phi_{BKS}$ are:
$\alpha=2.5$~nm$^{-1}$, $R_s=0.77476$~nm, $R_c=1$~nm, $q_{\rm
Si}=2.4e$ and $q_{\rm O}=1.2e$, where $e$ is the charge of an
electron.}
\label{tPAR}
\end{table}

\begin{table}
\begin{tabular}{|c|c|c|c|c|}
\hline
Label & $V$ (cm$^3$/mol) & $V$ (cm$^3$/g)
      & $\rho$ (g/cm$^3$)   &  L (nm)\\
\hline
A & 5.1359 & 0.256443 & 3.8995 & 2.2479818 \\
B & 5.6423 & 0.281722 & 3.5496 & 2.3195561 \\
C & 6.1486 & 0.307012 & 3.2572 & 2.3869665 \\
D & 6.6550 & 0.332292 & 3.0094 & 2.4507704 \\
E & 7.1614 & 0.357577 & 2.7966 & 2.5114146 \\
F & 7.6677 & 0.382863 & 2.6119 & 2.5692635 \\
G & 8.1741 & 0.408147 & 2.4501 & 2.6246184 \\
H & 8.4984 & 0.424340 & 2.3566 & 2.4157510 \\ 
I & 8.6804 & 0.433426 & 2.3072 & 2.6777320 \\
\hline
\end{tabular}
\caption{Isochore volumes, densities, and simulation box sizes studied
in this work.  All runs model $1332$ particles, except for isochore H,
where we model $999$ particles.  The reference volume $V_0$
corresponds to isochore I.}
\label{tISO}
\end{table}

\section{Configurational Entropy Calculation}
\label{predict}

In this section we calculate $S_c$ from knowledge of $e_{IS}$ and the
vibrational properties of the basins of the PES. Similar calculations
have been carried out for water~\cite{SCALA00}, binary LJ
mixtures~\cite{S01} and orthoterphenyl~\cite{MOSSA02}.

We begin by writing the Helmholtz free energy $F$ of the liquid along
an isochore as~\cite{SKT99},
\begin{equation}\label{eLIQFE}
F = e_{IS}(T) - TS_c(e_{IS}(T)) + f_{vib}(T,e_{IS}(T)).
\end{equation}
This expression separates $F$ into two contributions: one stemming
from the fact that the liquid samples different basins of the PES, and
one arising from the properties of the basins themselves.  $S_c$ is
the entropy contribution resulting from basin degeneracy, i.e. it
counts the number of basins associated with an inherent structure
energy $e_{IS}$~\cite{SKT99,S01}.  The vibrational part of the free
energy $f_{vib}$ arises from the free energy of the basins.  We note
that the basin properties may change with $e_{IS}$, e.g., the
vibrational density of states of a basin associated with low $e_{IS}$
may differ from that of one with high $e_{IS}$.  In equilibrium, $F$
is minimized with respect to $e_{IS}$ and we obtain
\begin{equation}\label{eqPRESCONF}
\frac{\partial{F}}{\partial{e_{IS}}}
= 0 = 1 - T\frac{\partial{S_c}}{\partial{e_{IS}}}
+ \frac{\partial{f_{vib}}}{\partial{e_{IS}}}.
\label{poof}
\end{equation}
We will provide evidence below that the vibrational properties of the
basins do not change substantially from one $IS$ to another.  In this
case $\partial f_{vib}/\partial e_{IS}=0$, and we may write,
\begin{equation}\label{eqSCONF}
S_c(T) = S_c(T_0)
+ \int_{T_0}^T \frac{1}{T'} \frac{\partial{e_{IS}}}{\partial{T'}} dT'.
\label{sss}
\end{equation}
$S_c(T_0)$ is the configurational entropy at a reference $T=T_0$.

Eq.~\ref{eqSCONF} shows that the behavior of $S_c(T)$ is controlled by
$e_{IS}(T)$.  The work of Sastry~\cite{S01} and
others~\cite{SKT99,starr} has shown that for fragile liquids, $e_{IS}$
decreases, and is negatively curved, as $T$ decreases.  In accordance
with Eq.~\ref{sss}, $S_c$ for fragile liquids also decreases, and is
negatively curved, as $T$ decreases.  If liquid silica is fragile at
high $T$, and then crosses over to strong behavior at lower $T$, we
expect that on cooling, both $e_{IS}$ and $S_c$ will initially behave
as in a fragile liquid.  However, from Eq.~\ref{eAG}, strong behavior
implies that $S_c$ (and hence $e_{IS}$) is constant with respect to
$T$.  Therefore, if the liquid passes from fragile to strong behavior
as $T$ decreases, the implication is that $e_{IS}(T)$ will decrease
with $T$ at high $T$, and then pass through a point of inflection,
consistent with the approach to a constant at low $T$.

To obtain $S_c(T_0)$ we write,
\begin{equation}
S_c(T_0)=S(T_0) - S_{vib}(T_0),
\label{sc}
\end{equation}
where $S(T_0)$ is the total liquid entropy and $S_{vib}(T_0)$ is the
entropy contribution arising from the vibrational properties of basins
in the PES.  The vibrational part has both a harmonic and an
anharmonic contribution, which we calculate separately:
\begin{equation}
S_{vib}(T) = S_{harm}(T) + S_{anh}(T),
\label{eqSVIB}
\end{equation}
where~\cite{PATHRIA}
\begin{equation}\label{eqSHARM}
S_{harm} = \frac{R}{N} \sum_{i=1}^{3N-3} \bigg( 1 - \ln{\frac{\hbar \omega_i}{k T}}\bigg),
\end{equation}
and
\begin{equation}\label{eqSANH}
S_{anh} = \int_0^T \frac{1}{T'} \frac{\partial{E_{anh}}}{\partial{T'}} dT'.
\end{equation}
Here $S_{harm}$ is the entropy in the harmonic approximation of the
IS's obtained at a given $T$.  The set $\{\omega_i\}$ describes the
vibrational density of states of the IS's (details below), $\hbar$ is
Planck's constant over 2$\pi$, $R$ is the gas constant, and $k$ is
Boltzmann's constant.  $E_{anh}$ is given by
\begin{equation}
\label{eqEANH}
E_{anh}(T) = E(T) - E_{harm}(T) - e_{IS}(T),
\end{equation}
where $E_{harm}$ is the harmonic contribution to the energy, given by
$E_{harm}=3RT(N-1)/N$.

To obtain $e_{IS}$ we select $100$ equilibrated liquid configurations
over the course of the MD run, perform a conjugate gradient
minimization~\cite{NR} of $\cal U$, and then average the results.  As
a stopping criterion for the conjugate gradient minimizations, we
specify a relative tolerance of $10^{-8}$ along line minimizations and
a relative tolerance of $10^{-15}$ between line minimizations.  In the
case of isochore H our runs are the longest, and so we average over
$1000$ configurations.

In order to evaluate $S_c$ and $S_{anh}$ (using Eqs.~\ref{eqSCONF} and
\ref{eqSANH} respectively) we first fit average values of $e_{IS}$ and
$E_{anh}$ to polynomials in $T$, and then evaluate the required
integrals analytically.  The $E_{anh}$ fit is constrained so that at
$T=0$, the value of $E_{anh}$ and its first derivative are zero.  This
is consistent with $E_{anh}$ being a correction to the harmonic
approximation.

It is important to recognize that the expression for $S_{anh}$ in
Eq.~\ref{eqSANH}, and hence the estimation of $S_c$ via
Eqs.~\ref{eqSCONF}-\ref{eqEANH}, is valid only under the assumption
that the basin anharmonicity does not change from basin to basin.  To
understand this, consider the expression for $E_{anh}$ in
Eq.~\ref{eqEANH}, from which $S_{anh}$ is calculated.  The terms
contributing to $E_{anh}$ are evaluated from equilibrium liquid
properties.  Yet, an implication of Eq.~\ref{eqEANH} is that an IS
obtained from a liquid at (e.g.) $4000$~K, when heated itself to
$4000$~K, will give a value of $E_{anh}$ equal to that obtained from
equilibrium configurations at $4000$~K.  However, an IS obtained from
a liquid at $3000$~K, when heated to $4000$~K will not necessarily
yield the value of $E_{anh}$ found from equilibrium configurations at
$4000$~K, because IS's obtained from different $T$ may be in basins of
different shape, and hence different anharmonicity.  If the basin
shape does change with $e_{IS}$, then $S_c$ in Eq.~\ref{eqSCONF} will
be influenced by an additional contribution.  Moreover,
Eq.~\ref{eqEANH} would be invalid and $E_{anh}$ would have to be
obtained in a different way, possibly by a careful heating of
individual basins obtained from the equilibrium liquid at different
$T$.  Such heating experiments must be performed with care, as the
system must not diffuse out of the basin if accurate results are to be
obtained.

\begin{figure}
\hbox to\hsize{\epsfxsize=1.0\hsize\hfil\epsfbox{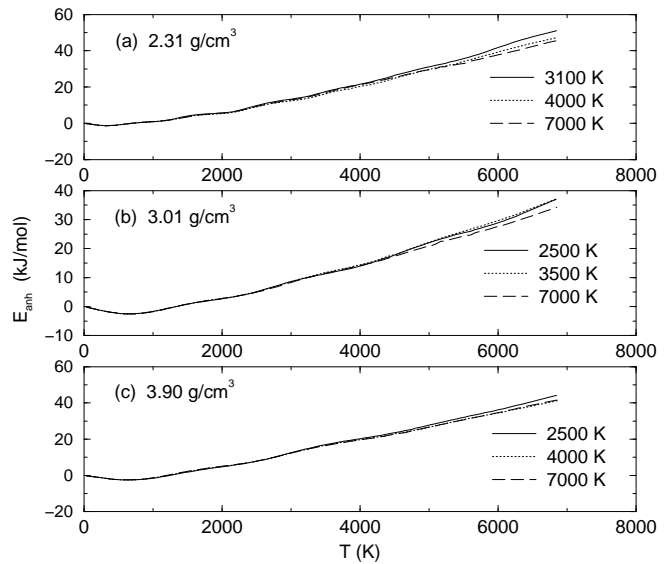}\hfil}
\caption{Test of $T$-dependence of basin shape.  IS's from three $T$,
and for three isochores, are rapidly heated in order to confine the
sampling to a single basin.  Using velocity scaling, $T$ is increased
from $0$~K to $7000$~K over $100$~fs.  Each curve is an average over
$10$ runs.  The curves for the same isochore are approximately the
same, indicating that the anharmonic contributions to the vibrational
energy can be assumed, for the present purposes, to be the same for
each basin}
\label{heat}
\end{figure}

To test if basins associated with different $T$ have different shapes,
we carry out runs in which IS's from different $T$ and $V$ are rapidly
heated.  We find that $E_{anh}$ is the same for all basins belonging
to the same isochore up to high $T$ (Fig.~\ref{heat}). 

Based on the relations justified above, we can evaluate $E_{anh}(T)$
and $S_{anh}(T)$ from a knowledge of $e_{IS}(T)$.  We can also
evaluate $S_c(T)$, up to a constant, from $e_{IS}(T)$.  To complete an
evaluation of $S_c(T)$, we need to estimate both $S_{harm}(T)$ and
$S(T_0)$ for each isochore to be studied, as described in the
following two subsections.


\subsection{Harmonic Entropy of Inherent Structures}

We define $S_{harm}$ of the liquid as the average harmonic entropy of
IS's sampled from the liquid.  When a liquid configuration is quenched
to its corresponding IS, it becomes a mechanically stable solid, and
is to a first approximation, harmonic.  To calculate the entropy of an
IS in the harmonic approximation, we require its vibrational density
of states.  As each IS is an atomic configuration at a local minimum
of the PES, we expand the expression for $\cal U$ about the local
minimum:
\begin{equation}
{\cal U} = e_{IS} + \sum_{i=1}^{3N} \sum_{j=1}^{3N} q_i
\frac{\partial^2{\cal U}}{\partial{q_i}\partial{q_j}}\bigg|_{q=q_0}
q_j.
\end{equation}
Here, the set $\{q_i\}$ specifies the 3N atomic coordinates, and the
notation ``$q=q_0$'' denotes that the second derivatives are evaluated
at the minimum energy configuration.  We then define a Hessian matrix,
\begin{equation}
H_{ij} = \frac{1}{\sqrt{m_i m_j}}
         \frac{\partial^2 {\cal U}}{\partial q_i \partial q_j}
         \biggr| _{q=q_0},
\label{hess}
\end{equation}
where $m_i$ is the mass of the atom associated with coordinate
$q_i$. Since the system is at a minimum, $H_{ij}$ has eigenvalues
$\{h_i\}$ all greater than zero, except for three zero eigenvalues
which account for the three independent translations of the entire
system. These three eigenvalues are excluded in calculating the
harmonic entropy. The ${w_i}$ appearing in Eq.~\ref{eqSHARM} are
defined as $w_i =\sqrt{h_i}$.  We note that a particular Hessian
matrix corresponds to an IS obtained from a liquid configuration at a
certain $T$.  It is this $T$ that we use in Eq.~\ref{eqSHARM}.

\begin{figure}
\hbox to\hsize{\epsfxsize=1.0\hsize\hfil\epsfbox{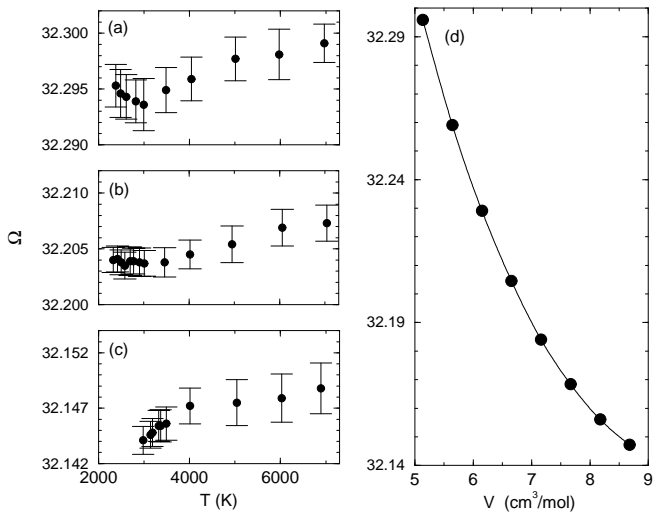}\hfil}
\caption{$\Omega$ as a function of $T$ for (a) isochore A, (b)
isochore D, and (c) isochore I.  Also shown is the standard deviation
about the mean value based on 100 samples. We note that a difference
in $\Omega$ of $0.01$ yields a change in entropy of $0.24$~J/mol~K.
For the purpose of this work, we therefore consider $\Omega$ to be
constant along isochores. (d) $\Omega(T=4000 \,{\rm K})$ as a function
of $V$; this is the value of $\Omega$ used in our calculations.}
\label{omega}
\end{figure}

We find, perhaps surprisingly, that the spectrum of $w_i$ does not
change appreciably with $T$ along isochores.  We plot in
Fig.~\ref{omega} the quantity,
\begin{equation}\label{eOMEGA}
\Omega(T) = \biggl\langle \frac{1}{3N-3} \sum_{i=1}^{3N-3} \ln{w_i}\biggr\rangle
\end{equation}
to show that the contribution to $S_c$ from changes in $\Omega$ as
$T$, and hence $e_{IS}$, is varied is negligible.  This quantity is
part of the expression for $S_{harm}(T)$.  It captures the average
quadratic shape of a basin and hence determines any dependence of
$S_{harm}$ on $e_{IS}$.  The plot shows $\Omega$ not to vary
appreciably with $T$, and we conclude that there is no contribution to
$S_c$ from $f_{vib}$, at least not from the harmonic portion.  

\begin{figure}
\hbox to\hsize{\epsfxsize=1.0\hsize\hfil\epsfbox{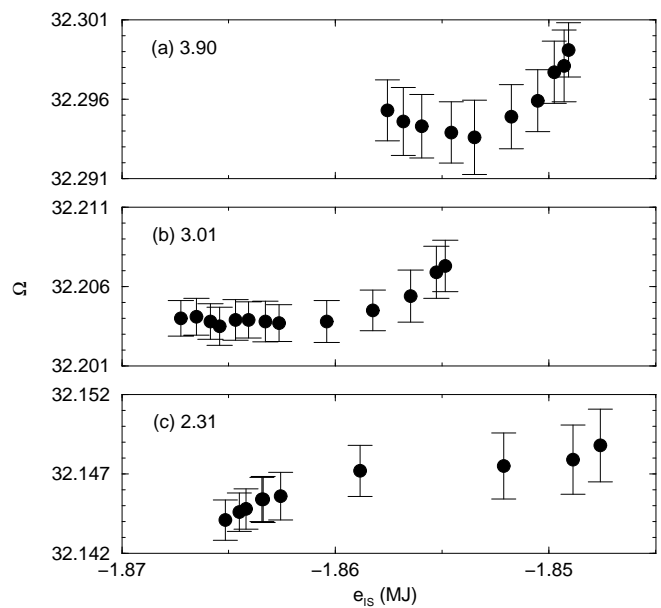}\hfil}
\caption{$\Omega$ as a function of $e_{IS}$ for (a) isochore A, (b)
isochore D, and (c) isochore I.  The error bars represent the standard
deviation about the mean value based on 100 samples.}
\label{omegaeis}
\end{figure}

To confirm this approximation, we show in Fig.~\ref{omegaeis} the
variation of $\Omega$ with $e_{IS}$.  We find that the change in
$\Omega$ can be as large as $\Delta\Omega\simeq 0.006$ for variations
of $e_{IS}$ as small as $\Delta e_{IS}\simeq 5$~kJ.  This gives a
contribution to $\partial f_{vib}/\partial e_{IS}$, the last term in
Eq.~\ref{poof}, of at most $RT\Delta\Omega / \Delta e_{IS}\simeq 0.04$.
This supports our assumption that $\partial f_{vib}/\partial
e_{IS}=0$.

\subsection{Liquid Entropy}

To exploit the AG relation, we require the absolute value of $S_c$,
not just changes in $S_c$ from one state point to another.  To
evaluate $S_c(T_0)$ in Eq.~\ref{sc} requires $S(T_0)$, the absolute
entropy of the BKS liquid, which we calculate via thermodynamic
integration starting from a system for which the entropy is known
exactly.

As our starting point we use the analytic result for the entropy of an
ideal gas composed of two species of particles, each with its own
mass~\cite{PATHRIA},
\begin{eqnarray}
S_{IG} &=& N_{Si} k \Biggl\{ \ln \Biggl[ \frac{V}{N_{Si}} \biggl(
         \frac{2 \pi m_{Si} k T}{h^2} \biggr)^{\frac{3}{2}} \Biggr]
         + \frac{5}{2} \Biggr\} \nonumber \\
       &+&
           N_O k \Biggl\{ \ln \Biggl[ \frac{V}{N_O} \biggl(
         \frac{2 \pi m_O k T}{h^2} \biggr)^{\frac{3}{2}} \Biggr]
         + \frac{5}{2} \Biggr\} \nonumber \\
         &-& k \ln\biggl(2\pi \sqrt{N_{Si} N_O}\biggr).
\label{sig}
\end{eqnarray}
For simplicity, we continue to label the two species as ``Si'' and
``O''.  Note that $h$ is Planck's constant, and that the Stirling
approximation has been employed in the derivation of this result, i.e.
$ \ln N! \approx N\ln N - N + \ln (2 \pi N)^{1/2}$.

For the purpose of thermodynamic integration, we approximate this
ideal gas with a dilute binary LJ system in which the stoichiometry of
the species is the same as that of Si and O in our silica simulations,
i.e. $N_{Si}=444$ and $N_{O}=888$.  

\begin{figure}
\hbox to\hsize{\epsfxsize=1.0\hsize\hfil\epsfbox{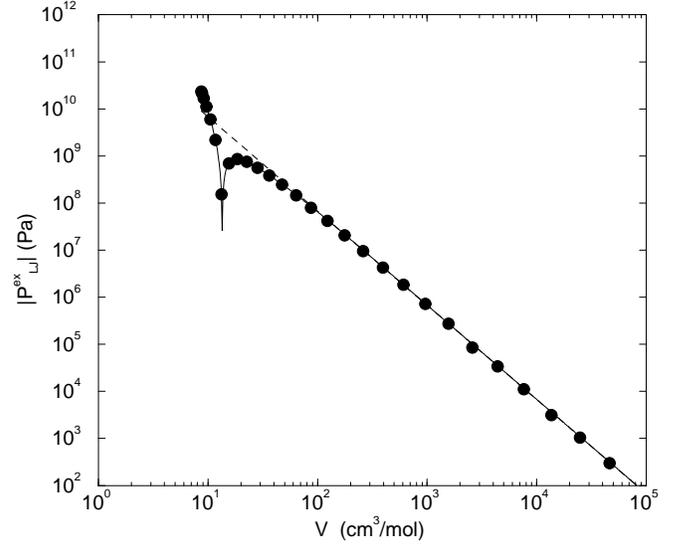}\hfil}
\caption{Isotherm of $|P^{ex}_{LJ}|$ for the LJ fluid at $T=4000$~K.
The solid line is the fit to Eq.~\protect{\ref{fit}} with $M=8$.  The
dashed line is the curve given by Eq.~\protect{\ref{pex}}.  The data
are shown on a log-log plot to simplify the comparison of the data to
Eq.~\protect{\ref{pex}} at large $V$.  The cusp near $V=10\, {\rm
cm^3/mol}$ is due to the fact that $P^{ex}_{LJ}$ changes sign.}
\label{compress}
\end{figure}

We equilibrate the LJ system at a reference $T=T_0=4000$~K over a
range of $V$ from a reference $V=V_0=8.6804 \, {\rm cm^3/mol}$ to
$V=173610\, {\rm cm^3/mol}$.  Denoting the reference state point at
($T_0,V_0$) as ``$C$'', the entropy of the LJ liquid $S_{LJ}$ at $C$
can be written,
\begin{equation}
S_{LJ}(C) = S_{IG}(C) + \frac{U_{LJ}(C)}{T} - \frac{1}{T}
           \int^\infty_{V_0} P^{ex}_{LJ} dV.
\label{slj}
\end{equation}
$S_{IG}(C)=117.236 {\rm J/mol~K}$ is calculated from Eq.~\ref{sig}.
From simulations we obtain $U_{LJ}(C)=-134099 \pm 50 \, {\rm J/mol}$,
the potential energy of the LJ system at $C$.  $P^{ex}_{LJ} = P_{LJ} -
NkT/V$ is the excess pressure of the LJ system, which we evaluate for
many $V$ at $T_0$ (Fig.~\ref{compress}).

In order to evaluate the integral in Eq.~\ref{slj}, we seek a function
to fit to our $P^{ex}_{LJ}$ data that we can integrate analytically.
A natural choice is the virial expansion for $P^{ex}_{LJ}$, a power
series in $1/V$.  At sufficiently large $V$, the $V$ dependence of
$P^{ex}_{LJ}$ will be well approximated by the leading term in the
virial expansion,
\begin{equation}
P^{ex}_{LJ} \simeq \frac{b_2 k T N^2}{V^2},
\label{pex} 
\end{equation}
where for our binary system,
\begin{equation}
 b_2 = 1/9 \,b_2^{SiSi} + 4/9\, b_2^{SiO} + 4/9 \,b_2^{OO},
\end{equation}
and the coefficients $b_2^{\mu\nu}$ are defined by,
\begin{equation}
b_2^{\mu\nu} = -4 \pi \int_0^\infty r^2 
\bigl(e^{-\Phi_{LJ}(\mu,\nu,r)/kT} -1\bigr) dr.
\end{equation}
We calculate $b_2$ numerically for the LJ system and find
$b_2=-0.0337~{\rm nm^3}$.  As shown in Fig.~\ref{compress}, we have
simulated the LJ fluid to large enough $V$ so that $P^{ex}_{LJ}$
conforms to Eq.~\ref{pex}.  To integrate $P^{ex}_{LJ}$ over
$(V_0,\infty)$, we fit the data to
\begin{equation}
P^{ex}_{LJ}=\frac{b_2 k T N^2}{V^2} + \sum_{n=3}^M \frac{a_n}{V^n},
\label{fit}
\end{equation}
and use this form to evaluate the integral in Eq.~\ref{slj}.  We
evaluate the integral using three different fits with $M=6$, $7$ and
$8$ in order to obtain an error estimate for the integral.  We
estimate the value of the integral to be $-1268 \pm 700 \, {\rm
J/mol}$, and thus find $S_{LJ}(C) = 84.028 \pm 0.175 $~J/mol~K.

\begin{figure}
\hbox to\hsize{\epsfxsize=1.0\hsize\hfil\epsfbox{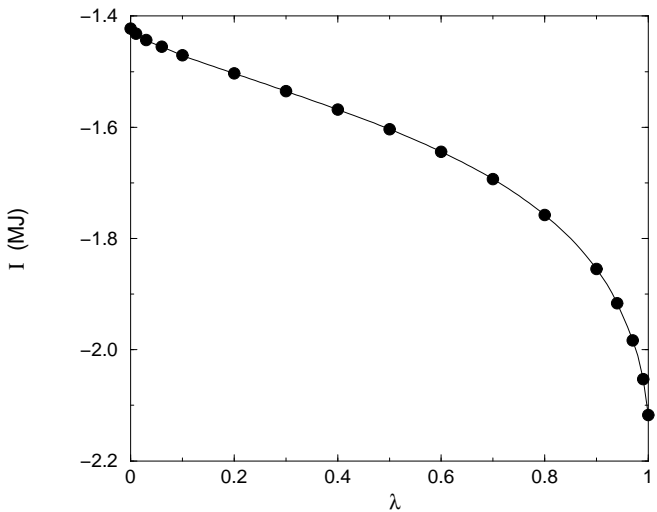}\hfil}
\caption{Plot of the integrand $I=\langle {\cal U}_{BKS} - {\cal
U}_{LJ}\rangle$, in Eq.{\protect{\ref{lambda}}}.  The solid line is a
cubic spline fit to the data.}
\label{morph}
\end{figure}

To obtain $S(C)$ from $S_{LJ}(C)$ we perform a generalized
thermodynamic integration~\cite{mezei}, in which a parameter $\lambda$
is used to create a continuous path between the LJ system at $C$ and
the BKS system at $C$.  To this end, we conduct MD simulations of a
system of particles interacting via a pair potential $\Phi$ such that,
\begin{equation}
\Phi(\lambda) = \lambda \Phi_{BKS} + (1-\lambda)\Phi_{LJ}.
\label{hybrid}
\end{equation}
When $\lambda=0$, the system corresponds to the LJ fluid, and when
$\lambda=1$, the system corresponds to the BKS potential.  For
arbitrary $\lambda$, the instantaneous potential energy is given by,
\begin{equation}
{\cal U}_\lambda = \lambda {\cal U}_{BKS} + (1-\lambda){\cal U}_{LJ},
\end{equation}
where ${\cal U}_{BKS}$ (${\cal U}_{LJ}$) is the instantaneous
potential energy of the system evaluated using only the $\Phi_{BKS}$
($\Phi_{LJ}$) pair potential.  The Helmholtz free energy difference
$\Delta F=F_{BKS}-F_{LJ}$ between the BKS and LJ systems at $C$ is
given by,
\begin{eqnarray}
\Delta F (C)
&=& \int_0^1
\Biggl \langle
\frac{\partial{{\cal U}_\lambda}}{\partial \lambda}
\Biggr \rangle_\lambda
d \lambda \nonumber \\
&=& \int_0^1 \langle {\cal U}_{BKS} - {\cal U}_{LJ}\rangle
d \lambda.
\nonumber \\
\label{lambda}
\end{eqnarray}
We evaluate the above integral by simulating the system governed by
Eq.~\ref{hybrid} at several values of $\lambda$ and using a cubic
spline to interpolate between points.  The integrand is shown in
Fig. \ref{morph}, from which we obtain $\Delta F (C)=-1635990 \pm 50
\, {\rm J}$ via numerical integration.

From our BKS simulations we find $U(C) = -1802257 \pm 100 \, {\rm
J/mol}$, from which we evaluate $\Delta U(C)=U(C)-U_{LJ}(C)$.  Using
$S(C)=S_{LJ}(C)+[\Delta U (C)-\Delta F (C)]/T$, we find for the BKS
liquid $S(C)=S(T_0,V_0)=75.986 \pm 0.177$~J/mol~K.

\begin{figure}
\hbox to\hsize{\epsfxsize=1.0\hsize\hfil\epsfbox{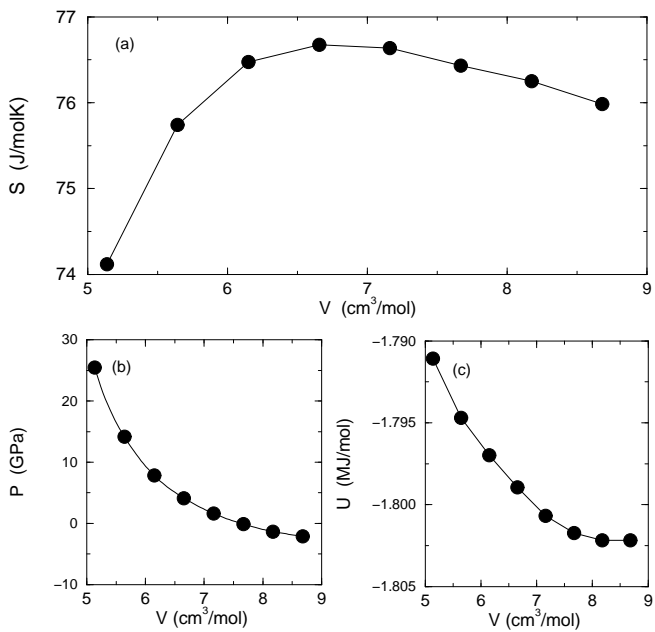}\hfil}
\caption{Thermodynamic properties of liquid BKS silica along the
$T_0=4000$~K isotherm: (a) $S$, (b) $P$, and (c) $U$.}
\label{ref}
\end{figure}

To obtain values of $S$ at $T_0$ for $V$ different from $V_0$, we
carry out a thermodynamic integration of $P$ along an isotherm of the
BKS liquid, using,
\begin{eqnarray}
S(V,T_0) 
&=& S(C) + \frac{1}{T}\bigl[U(V,T_0) - U(C)\bigr] \nonumber \\ 
&+& \frac{1}{T} \int_{V_0}^V P(V') d V'.
\end{eqnarray}
To evaluate the above integral, we find $P$ for various $V$ at $T_0$
and fit the data with a cubic spline.  This spline fit is then used to
generate data for a numerical evaluation of the integral.  We thus
have $S(V,T_0)$, the absolute entropy of the BKS liquid at all $V$
studied, at $T_0=4000$~K (Fig.~\ref{ref}).

\subsection{Crystalline Ground States}

As discussed in the next section, we also find it useful to calculate
the $T=0$ crystalline ground state energy $U(0)$ of the BKS system,
for comparison with the IS energies obtained from the quenched liquid
configurations.

\begin{figure}
\hbox
to\hsize{\epsfxsize=1.0\hsize\hfil\epsfbox{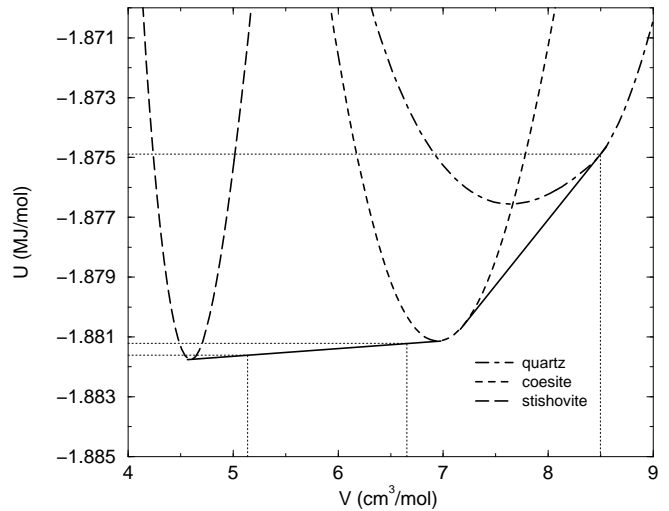}\hfil}
\caption{$U(V,T=0)$ for quartz, coesite and stishovite.  Solid lines
are common tangent constructions used to find the system ground state
energy for specific values of $V$ (as indicated by dotted lines).}
\label{fort102_2}
\end{figure}

To this end, we study three crystalline structures of silica important
in the $V$ range under consideration, namely quartz, coesite and
stishovite~\cite{SILICABOOK,chelik,smyth}. We evaluate the $T=0$
energy curves for these crystals as modeled by the $\Phi_{BKS}$ pair
potential.  Starting from the previously determined crystal
structures, we optimize $U(0)$ of the model system through an
iterative procedure where we alternately minimize $U(0)$ as a function
of the particle coordinates in a simulation cell of fixed geometry,
using a conjugate-gradient procedure; and then optimize the cell
geometry with a simplex method~\cite{NR}.  During the cell geometry
optimization, we constrain $V$ to be fixed, but otherwise allow the
shape to change.  This is done to remove anisotropic stress within the
crystal while preserving $V$. Once the crystal structure has been
optimized for a particular $V$, we incrementally change $V$ and repeat
the optimization. The results for the three crystals are shown in
Fig.~\ref{fort102_2}.  At fixed $V$, the thermodynamic ground state
may be a single crystal phase, or a coexisting mixture of two
crystalline phases.  To obtain the ground state energy in the case of
a mixture, we employ the ``common tangent construction,'' as shown in
Fig.~\ref{fort102_2}.

\section{Results and Discussion}

The calculations described above yield a complete thermodynamic
description of the BKS model of liquid silica, including the absolute
free energy of the model, and the absolute configurational entropy,
over a wide range of $V$-$T$ conditions.  Combined with dynamical
data, in the form of the diffusion coefficient $D$, a number of
conclusions may be drawn, as described below.  Note in the following
that by $D$ we mean the diffusion coefficient of the Si atoms,
evaluated from the particle mean squared displacement in equilibrium.
For comparison we also show some results for the diffusion coefficient
of the O atoms, $D_O$.  In general, the qualitative results are
independent of the choice of atomic species.

\subsection{Signature of fragile-to-strong crossover in the 
potential energy landscape}

\begin{figure}
\hbox to\hsize{\epsfxsize=1.0\hsize\hfil\epsfbox{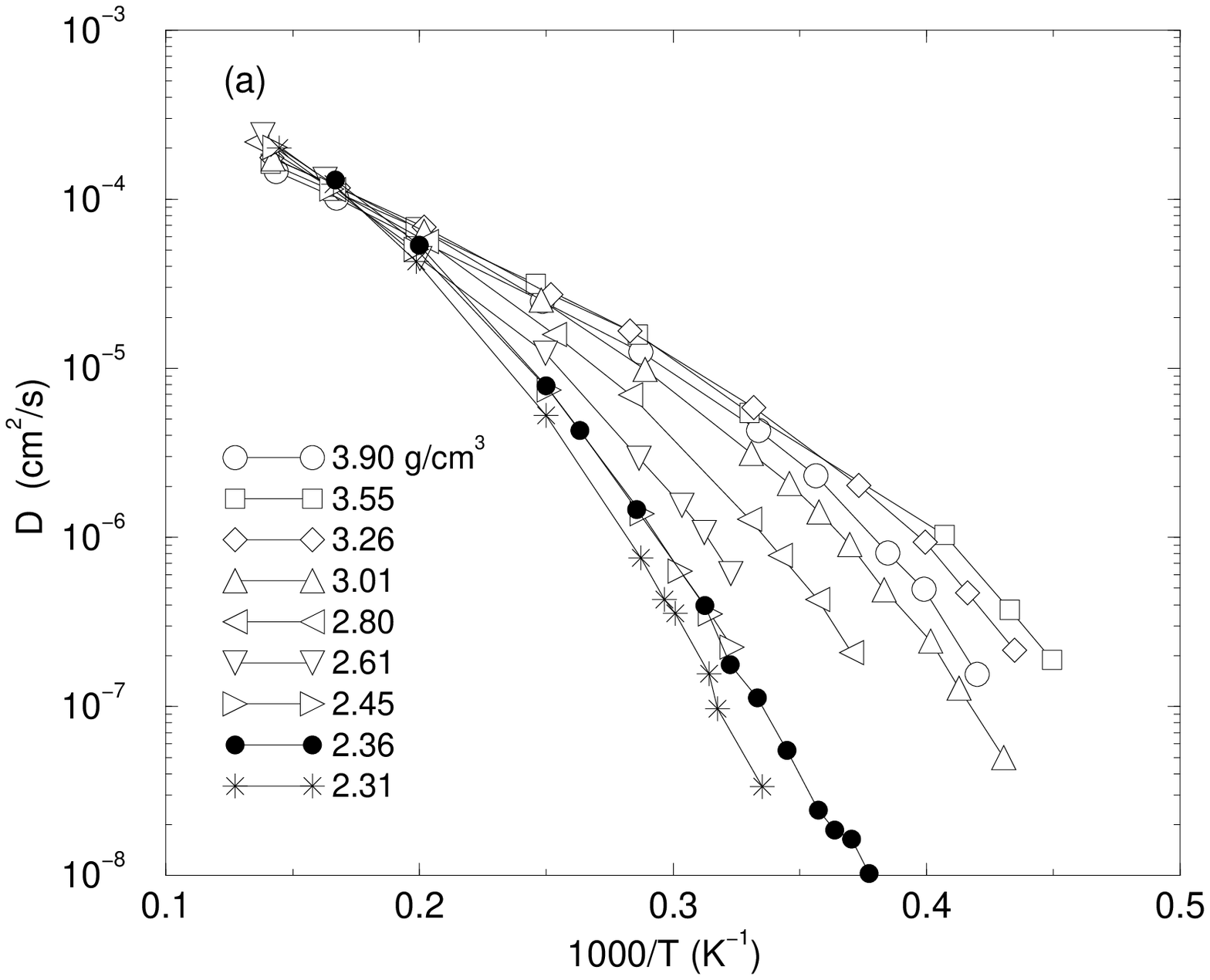}\hfil}
\hbox to\hsize{\epsfxsize=1.0\hsize\hfil\epsfbox{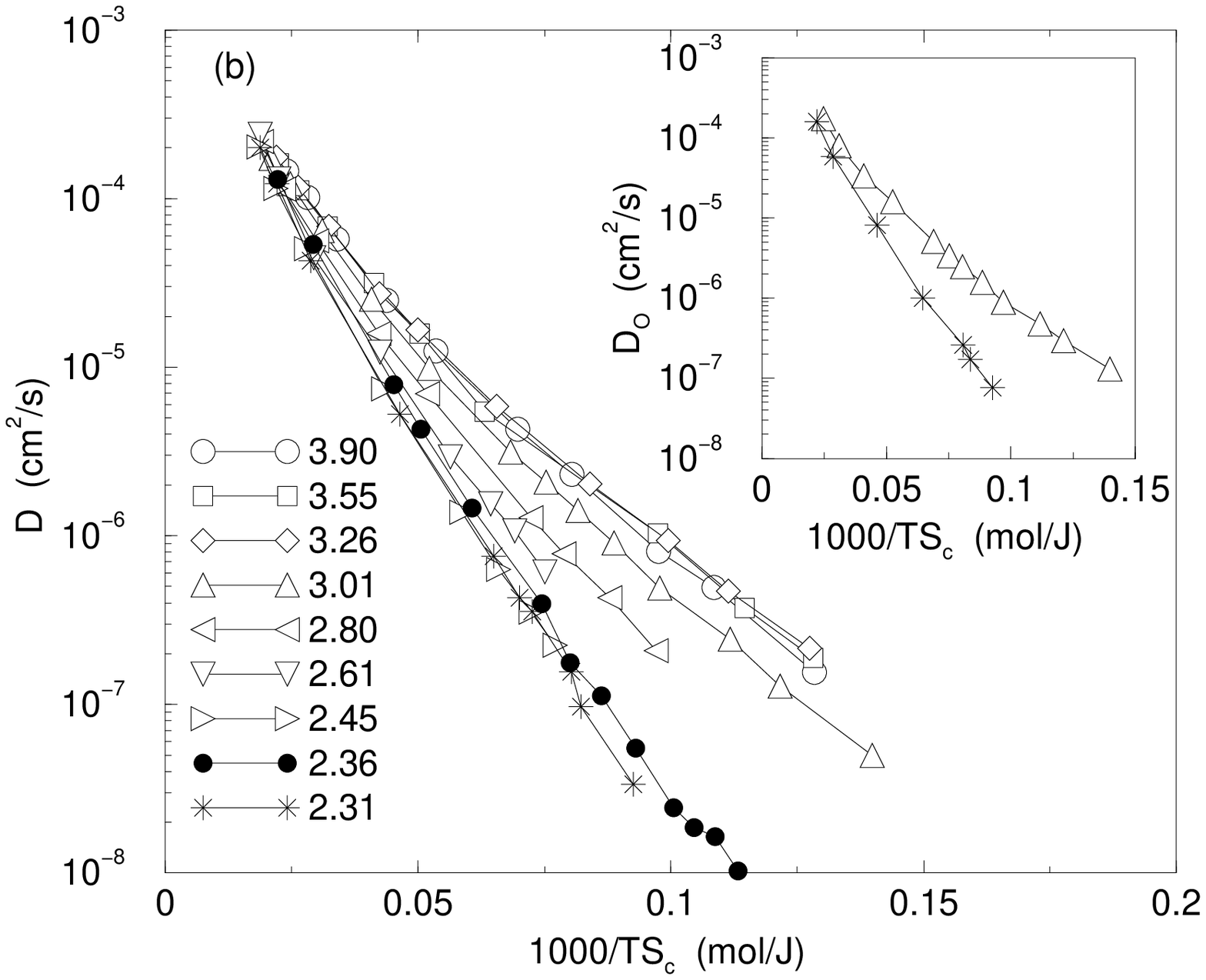}\hfil}
\caption{(a) Isochores of $D$, the diffusion coefficient of Si atoms.
(b) Test of the AG relation along isochores of $D$; the legend
indicates $\rho$ in g/cm$^3$.  If the data fall on a straight line,
the AG relation is satisfied.  For comparison, the inset shows $D_O$,
the diffusion coefficient of O atoms, along the $\rho=3.01$
(triangles) $\rho=2.31$~g/cm$^3$ isochores (stars).}
\label{diffd}
\end{figure}

\begin{figure}
\hbox to\hsize{\epsfxsize=1.0\hsize\hfil\epsfbox{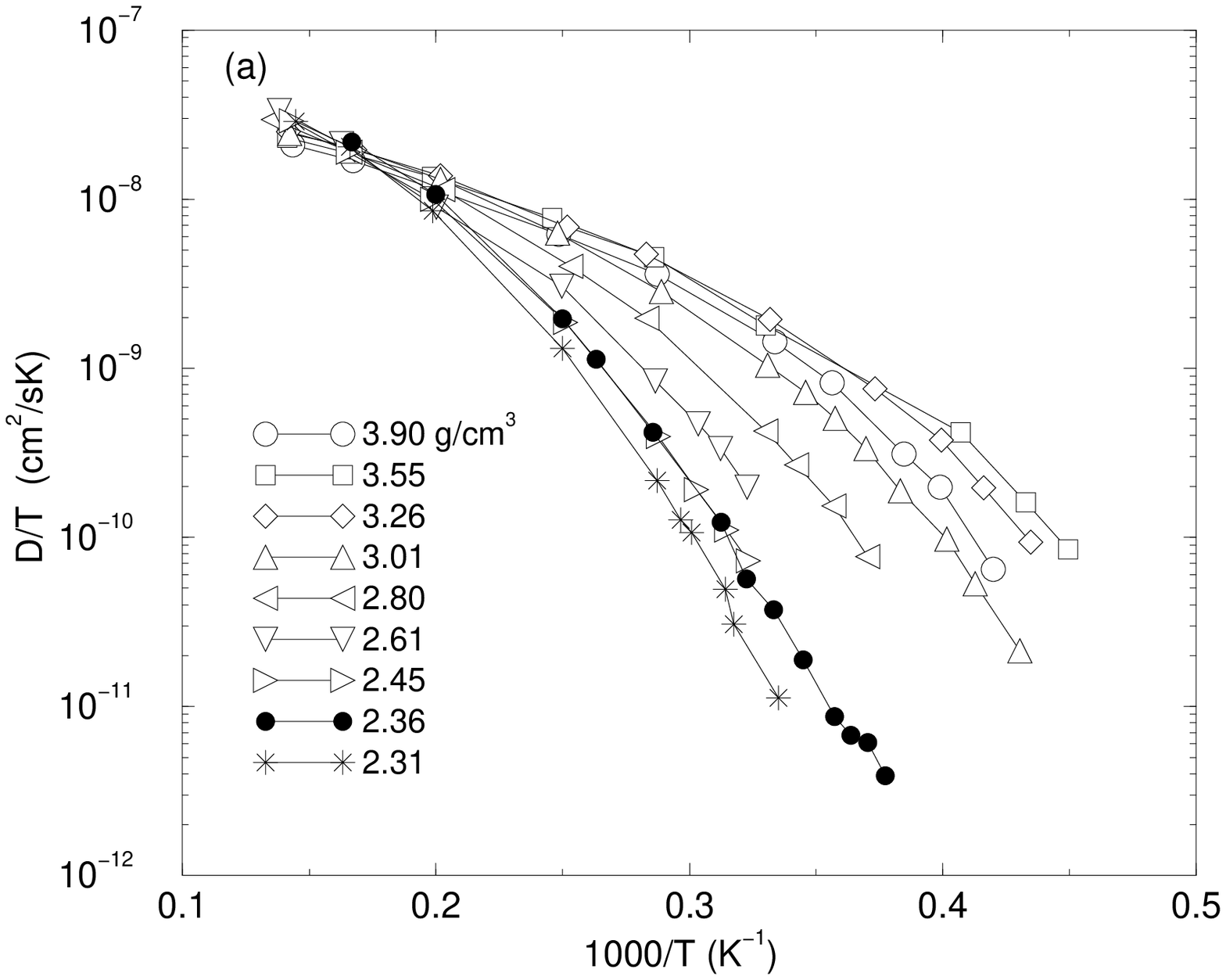}\hfil}
\hbox to\hsize{\epsfxsize=1.0\hsize\hfil\epsfbox{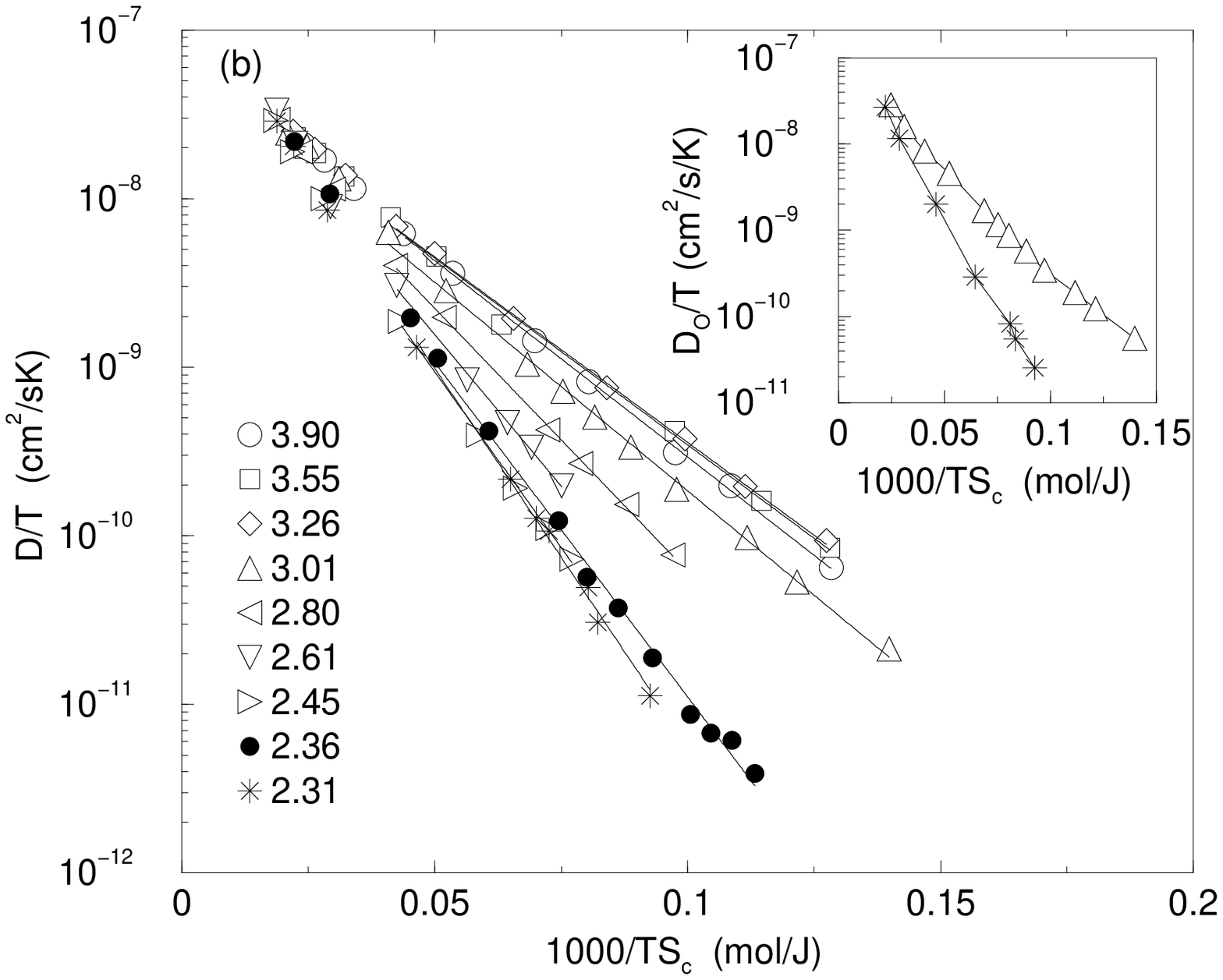}\hfil}
\caption{(a) Isochores of $D/T$.  (b) Test of the AG relation along
isochores of $D/T$; the legend indicates $\rho$ in g/cm$^3$.  For
comparison, the inset shows $D_O$ along the $\rho=3.01$ (triangles)
$\rho=2.31$~g/cm$^3$ isochores (stars).  The lines in the main panel
are fits of a straight line to the data, used to obtained the values
of $A$ and $\mu_0$ shown in Fig.~\protect{\ref{params-ag}}}.
\label{diffdt}
\end{figure}

Our results for $D$ are shown in Figs.~\ref{diffd}(a) and
\ref{diffdt}(a). These plots take the form of Arrhenius plots of $D$
and $D/T$ respectively, the latter quantity being preferred in some
works as a measure of particle mobility in liquids.  As first observed
in Ref.~\cite{HK99}, the FSC of the BKS model can be seen in our
simulations at low $\rho$, which correspond to the value of $\rho$ for
real silica at ambient $P$; isochores are curved on an Arrhenius plot
at high $T$, but becomes straighter at the lowest $T$.  

\begin{figure}
\hbox to\hsize{\epsfxsize=1.0\hsize\hfil\epsfbox{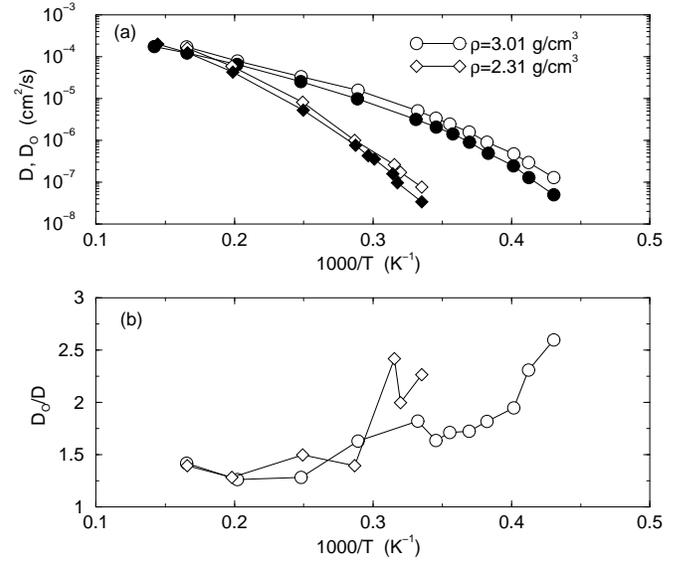}\hfil}
\caption{(a) Arrhenius plot of $D_O$ (open symbols) and $D$ (filled
symbols) along the $\rho=3.01$ (circles) and $\rho=2.31$~g/cm$^3$
(diamonds) isochores.  (b) Ratio $D_O/D$ as a function of $1/T$ along
the isochores shown in (a).}
\label{diffoxy}
\end{figure}

We compare the behavior of $D$ and $D_O$ in Fig.~\ref{diffoxy}.  We
find that the ratio of $D_O/D$ varies between $1$ and $3$, but that
the $T$ dependence of $D_O$ is qualitatively the same as that of $D$.

\begin{figure}
\hbox to\hsize{\epsfxsize=1.0\hsize\hfil\epsfbox{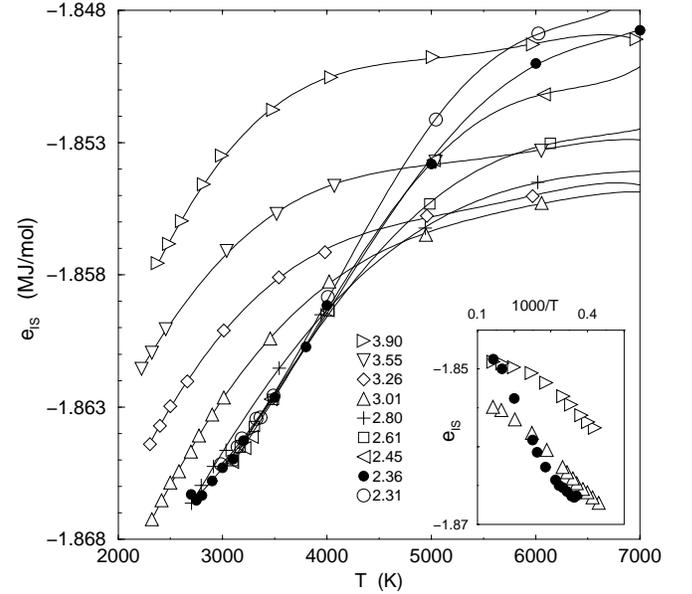}\hfil}
\caption{$e_{IS}(T)$ along isochores.  The lines are fits to each
isochore of a fifth order polynomial in $T$ with no linear term
(i.e. zero slope at $T=0$); these are the polynomial curves we use to
estimate the integral term in Eq.~\ref{sss}.  The legend indicates
$\rho$ in g/cm$^3$.  The inset shows $e_{IS}$ versus $1/T$ for three
isochores spanning the density range.  The low density isochore shows
a marked departure from the relation $e_{IS} \sim 1/T$ at low $T$.
The symbols used in the inset correspond to the same $\rho$ as in the
main panel.}
\label{eis3}
\end{figure}

\begin{figure}
\hbox to\hsize{\epsfxsize=1.0\hsize\hfil\epsfbox{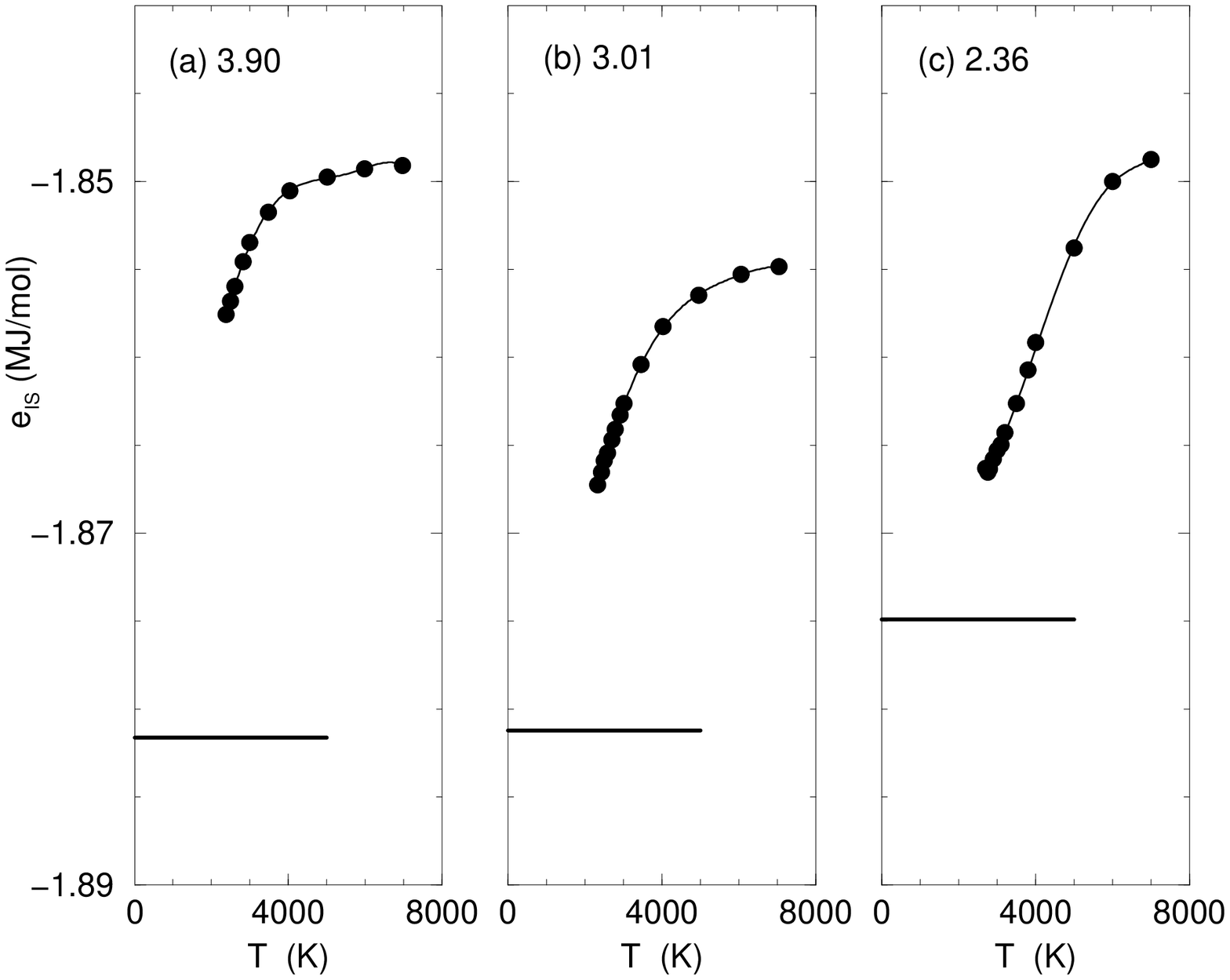}\hfil}
\caption{Detail of isochoric $e_{IS}$ behavior for three $\rho$
spanning the range of our calculations.  The thick horizontal lines
show the value of the $T=0$ crystal energies obtained from
Fig.~\protect{\ref{fort102_2}}.  }
\label{eis3b}
\end{figure}

Our results for $e_{IS}$ and $S_c$ are shown in Figs.~\ref{eis3},
\ref{eis3b} and \ref{sconf}.  Note that our estimates for $S_c$ have
changed somewhat, compared to the values published in
Ref.~\cite{NAT01}, due to improved averaging using more data, as well
as refinements in our analysis. However, the qualitative results of
Ref.~\cite{NAT01} remain in agreement with those presented here.

Consistent with the predictions made in Section~\ref{predict}, we find
that for the low $\rho$ isochores, where a FSC is observed, we also
observe a point of inflection in the $T$ dependence of both $e_{IS}$
and $S_c$.  This inflection is what one would expect if the emergence
of strong liquid behavior with decreasing $T$ is associated with the
approach of $e_{IS}$ and $S_c$ to a constant at low $T$.

Ref.~\cite{S01} showed that for a binary Lennard-Jones mixture at low
$T$, $e_{IS}(T) \sim -\frac{1}{T}$.  This $T$-dependence of $e_{IS}$,
consistent with a Gaussian distribution of IS energies, has been
observed in other models~\cite{starr,MOSSA02}.  Furthermore, the
binary LJ system is a relatively fragile liquid, and in this regard,
our results for silica at high $\rho$ are similar to those for binary
LJ (inset, Fig.~\ref{eis3}).  This is consistent with the fact that at
high $\rho$ the network structure of liquid silica is disrupted,
giving behavior more like that of simpler liquids, such as the binary
LJ system.  We also find that the $T$-dependence of $S_c$ becomes more
like that of a simple, fragile liquid (Fig.~\ref{sconf}) as $\rho$
increases.

\subsection{Implications for the Kauzmann paradox}

In Fig.~\ref{eis3b}, we compare the behavior of $e_{IS}$ to the $T=0$
energy of the crystalline state of the system, as found from the data
in Fig.~\ref{fort102_2}.  It is interesting to note how closely
$e_{IS}$ approaches the crystal energy at low $\rho$, compared to the
behavior at higher $\rho$.  We do not expect $e_{IS}$, the energy of a
disordered configuration obtained from the liquid at $T$, to ever be
less than that of the $T=0$ crystalline state of the system.  The
$T$-dependence of $e_{IS}$ at low $\rho$ is consistent with behavior
that would respect this constraint as $T\to 0$.  
At higher $\rho$, $e_{IS}$ does not approach the crystal energy as
closely as it does at lower $\rho$.  The conditions that might induce
an inflection in $e_{IS}$ are therefore not realized in the $T$ range
of our simulations.

\begin{figure}
\hbox to\hsize{\epsfxsize=1.0\hsize\hfil\epsfbox{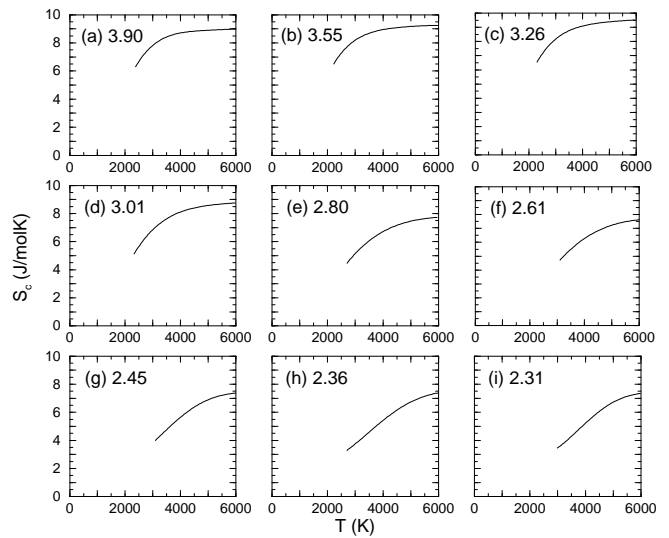}\hfil}
\caption{$S_c$ along isochores; each panel is labeled by the density
in g/cm$^3$.  Each curve is obtained using Eq.~\protect{\ref{sss}}, by
integrating the fitted curves for $e_{IS}$ shown in
Fig.~\protect{\ref{eis3}}.}
\label{sconf}
\end{figure}

The inflection in $e_{IS}$ is associated with an inflection in the
$T$-dependence of $S_c$, also found at low-$\rho$ (Fig.~\ref{sconf}).
For real systems, the third law of thermodynamics requires that the
lower bound for $S_c$ be zero.  Although our system is purely
classical, the same constraint applies, because the configurational
entropy we calculate counts the number of basins explored by the
liquid, which is necessarily one or greater.  As pointed out by
Kauzmann in 1948~\cite{kauz}, the entropy as $T$ decreases of many
supercooled liquids initially decreases at a sufficiently high rate so
as to suggest that the entropy might reach zero at finite $T$ (the
so-called ``entropy catastrophe'').  That this purely thermodynamic
event seems to be preempted by the occurrence of a kinetic event, the
glass transition, is the so-called ``Kauzmann paradox.''  At low
$\rho$ and high $T$, we find that $S_c$ behaves in similar fashion,
decreasing rapidly as $T$ decreases.  An extrapolation of the observed
high $T$ behavior raises the possibility that $S_c$ might reach zero
at finite $T$.  However, the inflection observed in the lower part of
our observed $T$ range establishes behavior that allows Kauzmann's
``entropy catastrophe'' to be avoided through a purely thermodynamic
phenomenon.

It is therefore tempting to speculate that our observations may be
transferable to other systems to which the Kauzmann paradox seems to
apply.  How $S_c>0$ is maintained in deeply supercooled liquids can
perhaps be understood in terms of the PES changes observed here.
Moreover, the PES change we find in BKS silica is correlated to the
fragile-to-strong dynamical crossover.  Hence it is possible that the
FSC and (the avoidance of) the Kauzmann paradox are fundamentally
interrelated phenomena.

However, if the above speculations are confirmed, it is important to
note the differences between silica and other supercooled liquids.
The picture developed above implies that the $T$ range of the
phenomenon by which silica avoids the Kauzmann paradox is above, and
widely separated from, $T_g$ in silica.  In other supercooled liquids,
the glass transition may occur at $T$ above, and thus obscure, the
PES changes found here.  More work on these possibilities is clearly
required.

\subsection{Test of the Adam-Gibbs relation}

In order to draw the conclusions given above, we must also test that
the liquid satisfies the AG relation.  If $S_c$ does not control
the behavior of $D$, then we will lack the basis required for making a
connection between the behavior of the PES and the liquid dynamics.

We perform this test by plotting $\log(D)$ [Fig.~\ref{diffd}(b)] and
$\log(D/T)$ [Fig.~\ref{diffdt}(b)] against $\log(1/TS_c)$.  The AG
relation is obeyed by data that follows a straight line on such a
plot.  We find that the isochores of $D/T$ provide the best agreement
with the AG relation.  Note also that both high and low $\rho$
isochores, regardless of whether $S_c$ exhibits an inflection, obey
the AG relation. Thus we see that regardless of dynamical regime
(fragile or strong), and regardless of inflections in the $T$
dependence of $S_c$ or $e_{IS}$, the liquid behaves so as to satisfy
the AG relation.  This observation reinforces the positive tests of
the AG prediction that have been documented in other work (see
e.g. Ref.~\cite{S01,SCALA00,MOSSA02,corezzi}).

\begin{figure}
\hbox to\hsize{\epsfxsize=1.0\hsize\hfil\epsfbox{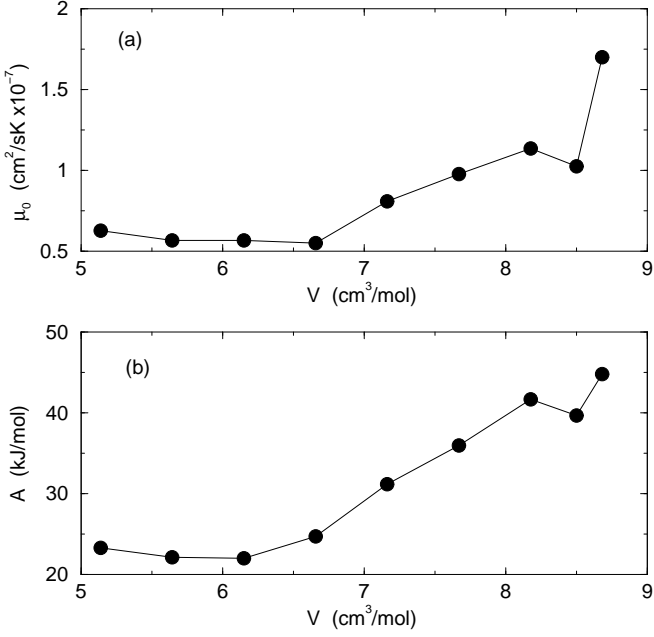}\hfil}
\caption{Estimates of the parameters (a) $\mu_0$ and (b) $A$, that
occur in Eq.~\protect{\ref{eAG}}.  These estimates are based on the
straight-line fits to the data shown in
Fig.~\protect{\ref{diffdt}}(b).}
\label{params-ag}
\end{figure}

In Fig.~\ref{params-ag} we present estimates of the constants $A$ and
$\mu_0$ that appear in Eq.~\ref{eAG}.  These are obtained by fitting
straight lines to the isochores in Fig.~\ref{diffdt}(b), omitting the
three data points at the highest $T$, where deviations from the AG
relation are expected.

\subsection{Entropy and Diffusion}

\begin{figure}
\hbox
to\hsize{\epsfxsize=1.0\hsize\hfil\epsfbox{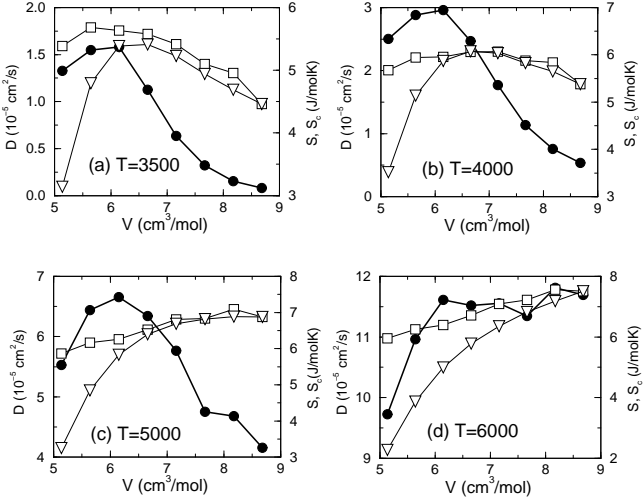}\hfil}
\caption{Isotherms of the $V$ dependence of $D$ (filled circles),
$S_c$ (squares) and $S$ (triangles) at various $T$.  For plotting
purposes, $S$ has been shifted down so that $S(V_0)=S_c(V_0)$. The
shifts in $S$ for the various panels are (a) $-66.6042 \, {\rm
J/mol~K}$, (b) $ -70.6031 \, {\rm J/mol~K}$, (c) $ -77.8919 \, {\rm
J/mol~K}$, and (d) $ -84.3673 \, {\rm J/mol~K}$.}
\label{ScDcomp}
\end{figure}

In the case of simulated water~\cite{SCALA00}, it was found that
maxima of $S_c$ isotherms occur, within error, at the same $V$ as the
maxima in isotherms of $D$.  We show our results for the $V$
dependence of $D$, $S_c$ and $S$ in Fig.~\ref{ScDcomp}.  At lower $T$,
isotherms of $S_c$ and $D$ pass through a maximum at approximately the
same $V$.  However, as $T$ increases, the correlation of these maxima
fades.  This may be due in part because at higher $T$, our estimates
of $S_c$ worsen, due to the larger role played by the anharmonic
corrections.  At the highest $T$, the trend is for isochores of $D$,
$S_c$ and $S$ to become monotonic functions of $V$.  An observation
that the $V$ dependence of entropy follows that of $D$ would be
consistent with recent work examining the relationship between
structural disorder and diffusivity in BKS silica~\cite{SHELL02}.

\subsection{Specific heat}

\begin{figure}
\hbox to\hsize{\epsfxsize=1.0\hsize\hfil\epsfbox{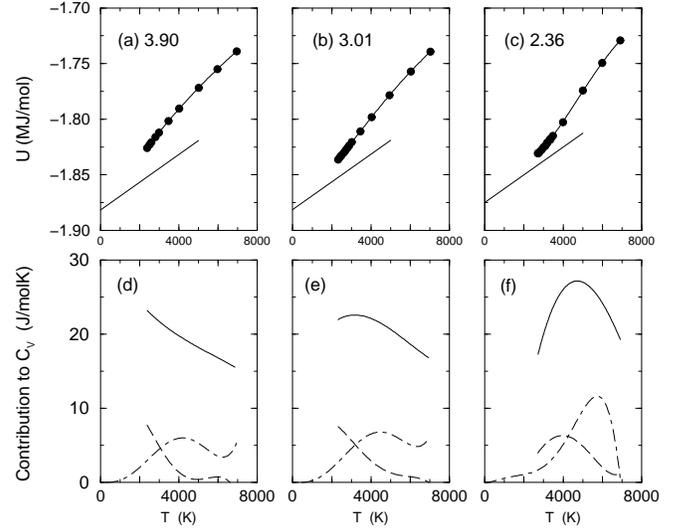}\hfil}
\caption{Comparison of $U$ and the contributions to $C_V$.  The top
panels (a)-(c) show isochores of $U$ spanning the studied density
range; also shown are estimates for $U$ of the crystalline state of
the system in the harmonic approximation, derived from the $T=0$
estimate of $U$ in Fig.~\protect{\ref{fort102_2}} and extended to
higher $T$ using a straight line of slope $3R/2$.  The bottom panels
(d)-(f) show the contributions to $C_V$.  The solid line is
$C_V-(3R/2)(N-1)/N$; the dot-dashed line is $C^{anh}_V$; and the
dashed line is $C^{IS}_V$.  Each curve is obtained by differentiating
the function fitted to the data for the corresponding energy.}
\label{ut2}
\end{figure}

In terms of the various contributions to
$E=e_{IS}+E_{anh}+3RT(N-1)/N$, we can write the constant volume
specific heat $C_V$ as,
\begin{eqnarray}
C_V &=& \biggl(\frac{\partial E}{\partial T}\biggr)_V \nonumber \\
    &=& \frac{\partial }{\partial T}\biggl |_V 
\biggl(e_{IS} + E_{anh} + \frac{3RT(N-1)}{N}\biggr).
\end{eqnarray}
So written, we can separately evaluate the contributions to $C_V$ from
$e_{IS}$ and $E_{anh}$, which we denote $C^{IS}_V$ and $C^{anh}_V$
respectively (Fig.~\ref{ut2}).  At all $\rho$, $C^{anh}_V$ exhibits a
maximum; at the lowest $\rho$ the inflection in the $T$ dependence of
$e_{IS}$ means that $C_V^{IS}$ also passes through a maximum.
Together, at low $\rho$, the strength of these two contributions
becomes large enough to give a peak in the total value of $C_V$.  This
peak is therefore a thermal signature approximately demarcating the
crossover from fragile to strong dynamical behavior.  It would be
interesting to explore if such signatures could be observed in high
$T$ experiments on silica, or related systems~\cite{bef2}.

\subsection{Relation to polyamorphism}

Real amorphous solid silica displays ``polyamorphism,'' the conversion
under pressure of a low density form to a high density form, that
occurs in some ways as though it were a first-order phase transition.
Computer simulations of BKS silica have provided evidence that this
polyamorphic transition may correspond to a sub-$T_g$ remnant of a
liquid-liquid phase transition occurring in the equilibrium
liquid~\cite{SAI01}.

Having found that the same model, BKS silica, exhibits a thermodynamic
anomaly, in the form of a $C_V$ peak associated with a FSC, it is
natural to ask if this phenomenon is related in some way to
polyamorphism.  It is difficult at present to answer this question
decisively, since the region of the proposed liquid-liquid instability
in BKS silica has only been approximately located, and seems to lie in
a $T$ range below that at which we can evaluate equilibrium liquid
properties with current computational resources.  However, several
trends suggest a connection.

\begin{figure}
\hbox
to\hsize{\epsfxsize=1.0\hsize\hfil\epsfbox{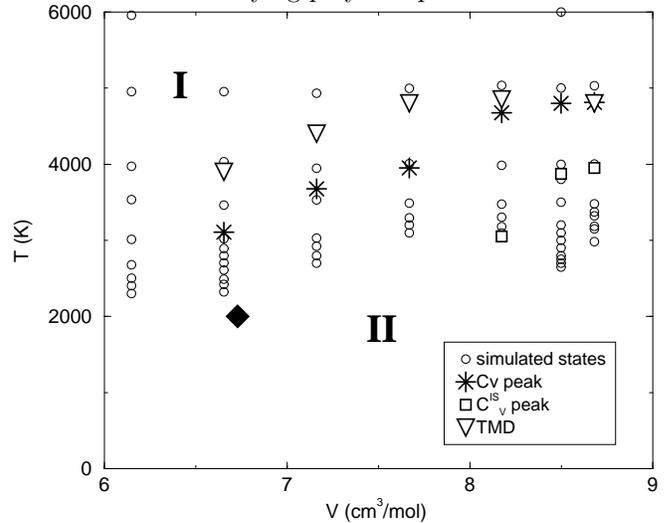}\hfil}
\caption{Location of the line of $C_V$ maxima (asterisks) in the
$V$-$T$ plane.  Also shown are points on the ``temperature of maximum
density'' (TMD) line (triangles), at which the isobaric expansivity
changes sign; and the location of maxima in the contribution of
$e_{IS}$ to $C_V$ (squares).  The diamond indicates where evidence for
liquid-liquid phase separation was found in Ref.~\cite{SAI01}.  The
regions labeled I and II are referred to in the text.}
\label{cv_peaks}
\end{figure}

First, we find that the $T$ at which the peak of $C_V$ occurs
decreases with increasing $\rho$, and passes outside of our range of
observation at the approximate $\rho$ where liquid-liquid phase
separation is proposed (Fig.~\ref{cv_peaks}).  This behavior is
consistent with the observed $C_V$ peaks being a line (in the $V$-$T$
plane) of high-$T$, non-singular thermodynamic anomalies that becomes
singular as the critical region of the liquid-liquid transition is
approached.

Second, the observed line of $C_V$ peaks naturally defines a
``crossover zone'' in the behavior of the liquid between a high-$T$,
high-$\rho$ region (Region I in Fig.~\ref{cv_peaks}), within which the
liquid is more fragile, the IS energies are relatively high, the
tetrahedral network is disrupted, and the properties are in general
more similar to simpler liquids; and a low-$T$, low-$\rho$ region
(Region II in Fig.~\ref{cv_peaks}) within which the liquid is becoming
strong, the IS energies are dropping (perhaps toward a lower limit),
the tetrahedral network is becoming prominent, and complex
thermodynamic behavior (e.g. negative expansivity) emerges.  It is
possible that these two regions of behavior, as $T$ decreases, become
progressively more sharply separated, perhaps ultimately by a
first-order phase transition.

More research, both through experiments and simulations, is required
to confirm or refute such a picture.  However, our current
understanding of the BKS system suggests that three distinct phenomena
may in fact be interrelated: (i) the FSC, (ii) polyamorphism, and
(iii) the landscape behavior that allows the Kauzmann paradox to be
avoided.  In this regard, Sasai has recently studied the
interrelationship of the FSC and a liquid-liquid phase transition in a
random energy model~\cite{sasai}.  Also, it is worth considering the
behavior of other systems (e.g. BeF$_2$~\cite{bef2}) that display one
or more of the above three phenomena, to test if the others also
appear. In particular, for any system that becomes strong at low $T$
via a FSC, it may be that polyamorphism can be observed under nearby
thermodynamic conditions.  This may be a useful clue for identifying
polyamorphic materials.

\section{Acknowledgments}
We thank C.A. Angell, W. Kob, S. Sastry and R. Speedy for discussions.
I.S.-V. and P.H.P thank NSERC (Canada) for funding and SHARCNET for
computing resources and funding. P.H.P. also acknowledges the support
of the Canada Research Chairs Program.  F.S. acknowledges support from
the INFM ``Iniziativa Calcolo Parallelo'' and PRA-HOP and from MIUR
FIRB and PRIN 2000.

\end{document}